\documentclass[aps,prl,twocolumn,superscriptaddress]{revtex4-1}
\usepackage[utf8]{inputenc}
\usepackage{amsmath}
\usepackage{amssymb}
\usepackage{epsfig}
\usepackage{graphicx,url}
\usepackage{bm}
\usepackage{mathrsfs}
\usepackage{tikz}
\usepackage{algcompatible}
\usepackage{newfloat}
\usepackage{caption}

\DeclareMathOperator*{\argmin}{argmin}

\newcommand{\redpathtwo}[2]{
\begin{tikzpicture}[baseline]
{\color{red} \draw[->] (0,0) to (0.25,0);
	\draw[->] (0.25,0) to (0.25+0.25*#1,0.25*#2);}
\end{tikzpicture}}

\newcommand{\redpaththree}[4]{
	\begin{tikzpicture}[baseline]
	{\color{red} \draw[->] (0,0) to (0.25,0);
		\draw[->] (0.25,0) to (0.25+0.25*#1,0.25*#2);
		\draw[->] (0.25+0.25*#1,0.25*#2) to (0.25+0.25*#1+0.25*#3,0.25*#2+0.25*#4);}
	\end{tikzpicture}}

\begin{document}

\title{Optimal non-markovian search strategies with $n$-step memory}

\author{Hugues Meyer}
\affiliation{Department of Theoretical Physics \& Center for Biophysics, Saarland University, 66123 Saarbrücken, Germany}
\author{Heiko Rieger}
\affiliation{Department of Theoretical Physics \& Center for Biophysics, Saarland University, 66123 Saarbrücken, Germany}

\date{\today}

\begin{abstract}
	Stochastic search processes are ubiquitous in nature and are expected to become more efficient when equipped with a memory, where the searcher has been before. A natural realization of a search process with long-lasting memory is a migrating cell that is repelled from the diffusive chemotactic signal that it secrets on its way, denoted as auto-chemotactic searcher. To analyze the efficiency of this class of non-Markovian search processes we present a general formalism that allows to compute the mean first passage time (MFPT) for a given set of conditional transition probabilities for non-Markovian random walks on a lattice. We show that the optimal choice of the $n$-step transition probabilities decreases the MFPT systematically and substantially with an increasing number of steps. It turns out that the optimal search strategies can be reduced to simple cycles defined by a small parameter set and that mirror-asymmetric walks are more efficient. For the auto-chemotactic searcher we show that an optimal coupling between the searcher and the chemical reduces the MFPT to 1/3 of the one for a Markovian random walk.
\end{abstract}

\maketitle

The term {\it search processes} encompasses all phenomena in which an agent scans a domain, looking for a target to reach. Search for preys and/or wild food resources by animals, known as foraging \cite{Pyke1977, Pyke1984, Stephens1986}, is one of the major examples of such processes. They can take various forms (blind or guided, individual or collective, random or deterministic, ...), but they all aim at being efficient, that is at minimizing the overall cost of the searching process. Several definitions for such a cost exist depending on the context, but it often simply reduces to the total duration of the search. In terms of statistical physics, the efficiency is usually quantified using first-passage time distributions: given all possible trajectories of the process considered, what is the probability that the agents will find the target in a certain amount of time? Optimizing the search efficiency therefore translates into minimizing of the first-passage time. The main statistical estimate is the mean first passage time (MFPT) although there are situations in which the whole FPT distribution is relevant \cite{Mattos2012, Godec2016}.

Many biological organisms, from bacteria to mammals, have evolved in such a way that their searching strategies are optimized in a certain way \cite{Smith1978}. Modeling these phenomena in quantitative terms is a challenge that has motivated many studies. Recently, various ways to transform simple blind random walks into efficient search processes have been suggested. Among other works, Bénichou and co-workers have e.g. shown that alternating periods of diffusive and ballistic motion can dramatically reduce first-passage times \cite{Benichou2005, Benichou2011}, and this strategy has actually been observed in various animal species. The effect of resetting on mean-first passage times and its efficiency as a search strategy have also been recently investigated \cite{Pal2017, Pal2020}. Other aspects such as the impact of confinement \cite{Condamin2005, Condamin2006} or the topology of the scanned domain \cite{Condamin2007}, have also been studied in different contexts. 

Memory of a stochastic process is also expected to affect the MFPT \cite{Hanggi1983, Hanggi1985, Masoliver1986, Verechtchaguina2006, Guerin2016}. A natural realization of a search process with a long-lasting memory is a migrating cell that is repelled from the diffusive chemotactic signal that it secrets on its way, denoted as auto-chemotactic searcher. Chemotaxis, a process in which a migrating cell changes its motion direction due to a chemical gradient of a chemical cue in its immediate surrounding, has been extensively studied, by biologists as well as chemists and physicists \cite{Adler1966, Keller1971, Macnab1972, jones2000, VanHaastert2004, Lammermann2013, Inoue2008, Iglesias2008, Stock2009, Taktikos2012, Lammermann2013, Pohl2014, Pezzotta2018}, and is, for instance, used by immune cells to guide themselves towards areas of infection or to tumors \cite{Krummel2016}. Experimental as well as theoretical studies of auto-chemotaxis are currently intensively studied in biophysics as it can help to understand the efficiency of a variety of biological processes \cite{Grima2005, Sengupta2009, Popescu2010, Zhao2013, Kranz2016, Jin2017, Stark2018}. Mathematically, these search processes are non-Markovian since the searcher uses the chemical information it has released along its past path in order to move ahead.

A systematic study of the first passage properties of stochastic non-Markovian search processes with $n$-step memory has not been performed yet, which is what we will present here: we will analyze the efficiency of non-Markovian search processes in general and of the auto-chemotactic in particular, and present results for the optimal search strategies. We first introduce a general formalism that allows to compute the MFPT for a given set of conditional transition probabilities for non-Markovian random walks on a lattice, which is based on the backward equation for the MFPT and on the conditional probability for the walker to go in a certain direction given its $n$ past directions. In the special case $n=1$ with additional constraints, we recover the results for the persistent random walk introduced in ref. \cite{Tejedor2012} by Tejedor and coworkers. Then we use this formalism to determine for a given $n$ the optimal conditional transition probabilities that minimize the MFPT. Finally we  analyze the search efficiency of the auto-chemotactic walker and determine the optimal coupling of the searcher to the self-generated chemotactic concentration field.

\vspace*{2ex}

\begin{figure}
	\begin{center}
		\includegraphics[width=.99\linewidth]{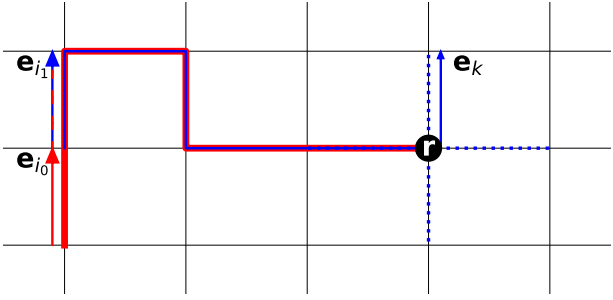}
		\caption{Sketch of a lattice walk illustrating the backward equation (\ref{eqn:backward_realtime}). The dotted lines indicate all possible sites the walker can jump to from its current position $\mathbf{r}$, given its past path $\mathbf{e}_{i_{0}}, \cdots, \mathbf{e}_{i_{n-1}}$}
		\label{fig:sketch_backward}
	\end{center}
\end{figure}
In this letter we consider a minimal model for a non-Markovian searcher: a random walk on a discrete lattice of lateral size $L$ and with transition probabilities depending on the steps it made before. Formally this stochastic process is defined by the  hierarchy of conditional transition probabilities $p\left(\mathbf{e}_{k} | \mathbf{e}_{i_{0}}, \cdots, \mathbf{e}_{i_{n-1}} \right)$, where $\mathbf{e}_{k}$ is the jump direction in the next step and $\left\{\mathbf{e}_{i_{0}}, \cdots, \mathbf{e}_{i_{n-1}}\right\}$ the jump directions of the last $n$ steps. This allows to write a backward equation of motion for the average first-passage time $T_n(\mathbf{r}, \mathbf{r}_T; \mathbf{e}_{i_{0}}, \cdots, \mathbf{e}_{i_{n-1}})$ to reach the target at position $\mathbf{r}_{T}$ for a walker starting at position $\mathbf{r}$ with the $n$ past directions $\left\{\mathbf{e}_{i_{0}}, \cdots, \mathbf{e}_{i_{n-1}}\right\}$:
\begin{align}
\label{eqn:backward_realtime}
T_n&(\mathbf{r}, \mathbf{r}_T; \mathbf{e}_{i_{0}}, \cdots, \mathbf{e}_{i_{n-1}}) = 1 +  \\
\sum_{k}& p\left(\mathbf{e}_{k} | \mathbf{e}_{i_{0}}, \cdots, \mathbf{e}_{i_{n-1}} \right)T(\mathbf{r}+\mathbf{e}_{k}, \mathbf{r}_T; \mathbf{e}_{i_{1}}, \cdots,\mathbf{e}_{i_{n-1}}, \mathbf{e}_{k}) \nonumber
\end{align}
The sum runs over all $z$ nearest neighbor sites the searcher can jump to, with $z$ the coordination number of the lattice. A sketch of equation (\ref{eqn:backward_realtime}) is shown in figure \ref{fig:sketch_backward} . We assumed periodic boundary conditions, which is equivalent to an infinite lattice with periodically arranged targets. In addition, equation (\ref{eqn:backward_realtime}) also holds for reflecting boundary conditions if one assumes that the target is placed at the center of the domain and that the probabilities  $p\left(\mathbf{e}_{k} | \mathbf{e}_{i_{0}}, \cdots, \mathbf{e}_{i_{n-1}} \right)$ are mirror-symmetric as we discuss in more detail below. Finally, eqn. (\ref{eqn:backward_realtime}) is obviously not correct if $\mathbf{r} = \mathbf{r}_{T}$, for which the average passage-time is trivially 0. In this case, the right-hand side yields the average return time on the site $\mathbf{r}_{T}$, equal to $V\equiv L^d$ \cite{Aldous1999}. By applying a discrete Fourier transformation $\tilde{f}(\mathbf{q}) = \sum_{\mathbf{r} \in \mathcal{L}} f(\mathbf{r}) e^{-i\mathbf{q}\cdot \mathbf{r}}$ with $q_{i} = 2\pi n_{i}/L$ and $n_{i}\in [\![ 0, L-1 ]\!]$, and properly accounting for the case $\mathbf{r} = \mathbf{r}_{T}$, a closed set of linearly coupled equations for $\tilde{T}_n(\mathbf{q}, \mathbf{r}_T; \mathbf{e}_{i_{0}}, \cdots, \mathbf{e}_{i_{n-1}})$ for all possible paths $\left\{\mathbf{e}_{i_{0}}, \cdots, \mathbf{e}_{i_{n-1}} \right\}$ is obtained, which can be cast into a matrix equation.

Let $\mathbf{s}_n$ be a vector of size $z^n$ containing all possible paths $\left\{\mathbf{e}_{i_{0}}, \cdots, \mathbf{e}_{i_{n-1}} \right\}$, and $\tilde{\mathbf{t}}_n$ a vector of equal size whose entries are defined as $\tilde{t}_{n\alpha}(\mathbf{q}, \mathbf{r}_T) = T_n(\mathbf{q}, \mathbf{r}_T; {s_n}_{\alpha})$. The solution of the matrix equation then is 
\begin{equation}
	 \tilde{\mathbf{t}}_n(\mathbf{q}, \mathbf{r}_T) = V\left[ \delta(\mathbf{q}) - e^{-i\mathbf{q}\cdot\mathbf{r}_{T}}\right] \left(\mathbb{I} - \mathbf{P}_n\mathbf{E}_n(\mathbf{q}) \right)^{-1} \mathbf{u}_n
\end{equation}
Here $\mathbf{u}_n$ is a vector of size $z^n$, all entries of which are equal to 1, $\mathbf{E}_n(\mathbf{q})$ is a square diagonal matrix whose elements are the complex exponentials $e^{i\mathbf{q}\cdot \mathbf{e}_k}$, and $\mathbf{P}_n$ is a square matrix containing all conditional probabilities $p\left(\mathbf{e}_{k} | \mathbf{e}_{i_{0}}, \cdots, \mathbf{e}_{i_{n-1}} \right)$. Note that this matrix has only $z^{n+1}$ non-zero elements, whose positions in the matrix depend on the ordering of the vector $\mathbf{s}_n$ \footnote{see Appendix A of supplemental material}. 

Fourier inversion and averaging over all possible initial positions yields 
\begin{equation}
\left\langle \mathbf{t}_n \right\rangle = \sum_{\mathbf{q}\neq 0} \left(\mathbb{I} - \mathbf{P}_n\mathbf{E}_n(\mathbf{q}) \right)^{-1} \mathbf{u}_n
\end{equation}
The mean first-passage time is finally computed by summing all entries of this averaged vector $\left\langle \mathbf{t} \right\rangle$, weighted by the probability of the respective paths. These weights are found using the identity
\begin{align}
	p&(\mathbf{e}_{i_{0}}, \cdots, \mathbf{e}_{i_{n-1}})  \\
	&=\sum_{i_{-1}} p(\mathbf{e}_{i_{n-1}} | \mathbf{e}_{i_{-1}}, \mathbf{e}_{i_{0}}, \cdots, \mathbf{e}_{i_{n-2}})p(\mathbf{e}_{i_{-1}}, \cdots, \mathbf{e}_{i_{n-2}}) \nonumber 
\end{align}
together with the normalization constraint $\sum_{i_{0},\cdots,i_{n-1}} p(\mathbf{e}_{i_{0}}, \cdots, \mathbf{e}_{i_{n-1}}) = 1$. These equations can again be cast into a matrix  form $\mathbf{M}_n\mathbf{p}_n = \mathbf{v}_n$. Here, $\mathbf{p}_n$ is a vector containing all entries of $p(\mathbf{e}_{i_{0}}, \cdots, \mathbf{e}_{i_{n-1}})$. $\mathbf{M}_n$ is equal to $\mathbb{I} - \mathbf{P}_n^{T}$ except for the last row, all elements of which are 1. Finally $\mathbf{v}_n$ is a vector containing only zeros except the last element being 1. The mean first-passage time is therefore obtained as the dot product $\left\langle T_n \right\rangle = \mathbf{p}_n\cdot\left\langle \mathbf{t}_n \right\rangle$.
This general formalism allows to infer the mean first-passage time of any non-Markovian random walk, provided the $n$-step conditional probability $p\left(\mathbf{e}_{k} | \mathbf{e}_{i_{0}}, \cdots, \mathbf{e}_{i_{n-1}} \right)$ is known \footnote{A discussion of the computational cost can be found in Appendix B of the supplemental material}. \newline

\begin{figure*}
	\begin{center}
		\begin{minipage}{0.48\linewidth}
			\begin{tabular}{|c|c|c|}
				\hline 
				$n=1$ & \multicolumn{2}{c|}{$p_{f}=q^{(1)}_0$, $p_{l}=p_{r}=\left(1-q^{(1)}_0\right)/2$, $p_{b}=0$}\\
				\hline
				& \centering	\underline{Asymmetric} & 	\underline{Symmetric} \\
				$n=2$ & 
				\begin{tikzpicture}[>=stealth,every node/.style={shape=rectangle,rounded corners},baseline={(0,-0.75)}]
				\node (c1) at (0,0) [draw]{\redpathtwo{0}{-1}};
				\node (c2) at (-1,-1) [draw] {\redpathtwo{1}{0}};
				\node (c3) at (1,-1) [draw] {\redpathtwo{0}{1}};
				\node (c4) at (-1,-1.7) [draw] {\redpathtwo{-1}{0}};
				\draw [->, blue, thick] (0,-0.9) arc [start angle=-90, end angle=260, x radius=0.5cm, y radius=0.2cm];
				\draw [<-, green, thick, rotate=-45, dashed] (0.7,0.2) arc [start angle=-90, end angle=260, x radius=0.4cm, y radius=0.2cm];
				\draw[->] (c1) to (c2);
				\draw[->] (c1) to[out=40,in=50] (c3) ;
				\draw[->] (c2) to (c3);
				\draw[->] (c4) to (c2);
				\draw[->] (c3) to  (c1);
				\draw (-0.8,0) node[below, rotate=45]{$p_0^{(2)}$} ;
				\end{tikzpicture} & 
				\begin{tikzpicture}[>=stealth,every node/.style={shape=rectangle,rounded corners}, baseline={(0,-0.5)}]
				\node (c1) at (0,0) [draw] {\redpathtwo{1}{0}} edge [loop above] node {} (); 
				\node (c2) at (1,-1) [draw] {\redpathtwo{0}{1}};
				\node (c3) at (-1,-1) [draw] {\redpathtwo{0}{-1}};
				\node (c4) at (0,-1.75) [draw] {\redpathtwo{-1}{0}};
				\draw[->] (c1) to (c3) ;
				\draw[->] (c3) to[out=90,in=180]  (c1) ;
				\draw[->] (c1) to (c2) ;
				\draw[->] (c2) to[out=90,in=0]  (c1) ;
				\draw[->] (c2) to[out=170,in=10]  (c3) ;
				\draw[->] (c3) to[out=350,in=190]  (c2) ;
				\draw[->] (c4) to (c2) ;
				\draw[->] (c4) to (c3) ;
				\draw (0.3,0.95) node[below]{$q_0^{(2)}$} ;
				\draw (0.05,-0.3) node[below]{$q_1^{(2)}$} ;
				\draw (0.6,-1.32) node[below]{$\frac{1}{2}$} ;
				\end{tikzpicture}\\
				\hline 
				$n=3$ &  \multicolumn{2}{c|}{ \begin{tikzpicture}[>=stealth,every node/.style={shape=rectangle,rounded corners},baseline={(0,-1)}]
					\node (c1) at (0,0) [draw] {\redpaththree{1}{0}{0}{1}};
					\node (c2) at (1.2,0) [draw] {\redpaththree{0}{1}{0}{1}};
					\node (c3) at (2.25,-0.25) [draw] {\redpaththree{1}{0}{1}{0}};
					\node (c4) at (2.25,-1.) [draw] {\redpaththree{1}{0}{0}{-1}};
					\node (c5) at (1,-1.5) [draw] {\redpaththree{0}{-1}{1}{0}};
					\node (c6) at (0,-1.5) [draw] {\redpaththree{0}{1}{-1}{0}};
					\node (c7) at (-1,-1.5) [draw] {\redpaththree{0}{1}{1}{0}};
					\node (c8) at (-1.75,-0.75) [draw] {\redpaththree{0}{-1}{-1}{0}};
					\node (c9) at (-1,0) [draw] {\redpaththree{0}{-1}{0}{-1}};
					\node (c10) at (2.25,0.5) [draw] {\redpaththree{-1}{0}{-1}{0}};
					\node (c11) at (-2.75,-0.75) [draw] {\redpaththree{-1}{0}{0}{1}};
					\node (c12) at (-2.75,0) [draw] {\redpaththree{1}{0}{-1}{0}};
					\node (c13) at (-2.75,-1.5) [draw] {\redpaththree{0}{-1}{0}{1}};
					\node (c14) at (-1,-2.25) [draw] {\redpaththree{-1}{0}{0}{-1}};
					\node (c15) at (0,-2.25) [draw] {\redpaththree{-1}{0}{1}{0}};
					\node (c16) at (1,-2.25) [draw] {\redpaththree{0}{1}{0}{-1}};
					\draw [<-, blue, thick] (0.2,-1.2) arc [start angle=-90, end angle=260, x radius=1.65cm, y radius=0.44cm];
					\draw [<-, green, thick, dashed] (-0.7,-1) arc [start angle=-90, end angle=260, x radius=0.6cm, y radius=0.25cm];
					\draw[->] (c1) to (c2);
					\draw[->] (c2) to (c3);
					\draw[->] (c3) to (c4);
					\draw[->] (c4) to (c5);
					\draw[->] (c5) to (c6);
					\draw[->] (c6) to (c7);
					\draw[->] (c7) to (c8);
					\draw[->] (c8) to (c9);
					\draw[->] (c9) to (c1);
					\draw[->] (c1) to (c6);
					\draw[->] (c10) to (c3);
					\draw[->] (c11) to (c8);
					\draw[->] (c12) to (c11);
					\draw[->] (c13) to (c11);
					\draw[->] (c14) to (c7);
					\draw[->] (c15) to (c14);
					\draw[->] (c16) to (c15);
					\draw (0.7,0) node[above]{$p_0^{(3)}$} ;
					\end{tikzpicture} } \\
				\hline
			\end{tabular}
			
		\end{minipage}
		\begin{minipage}{0.51\linewidth}
			\includegraphics[width=0.99\linewidth]{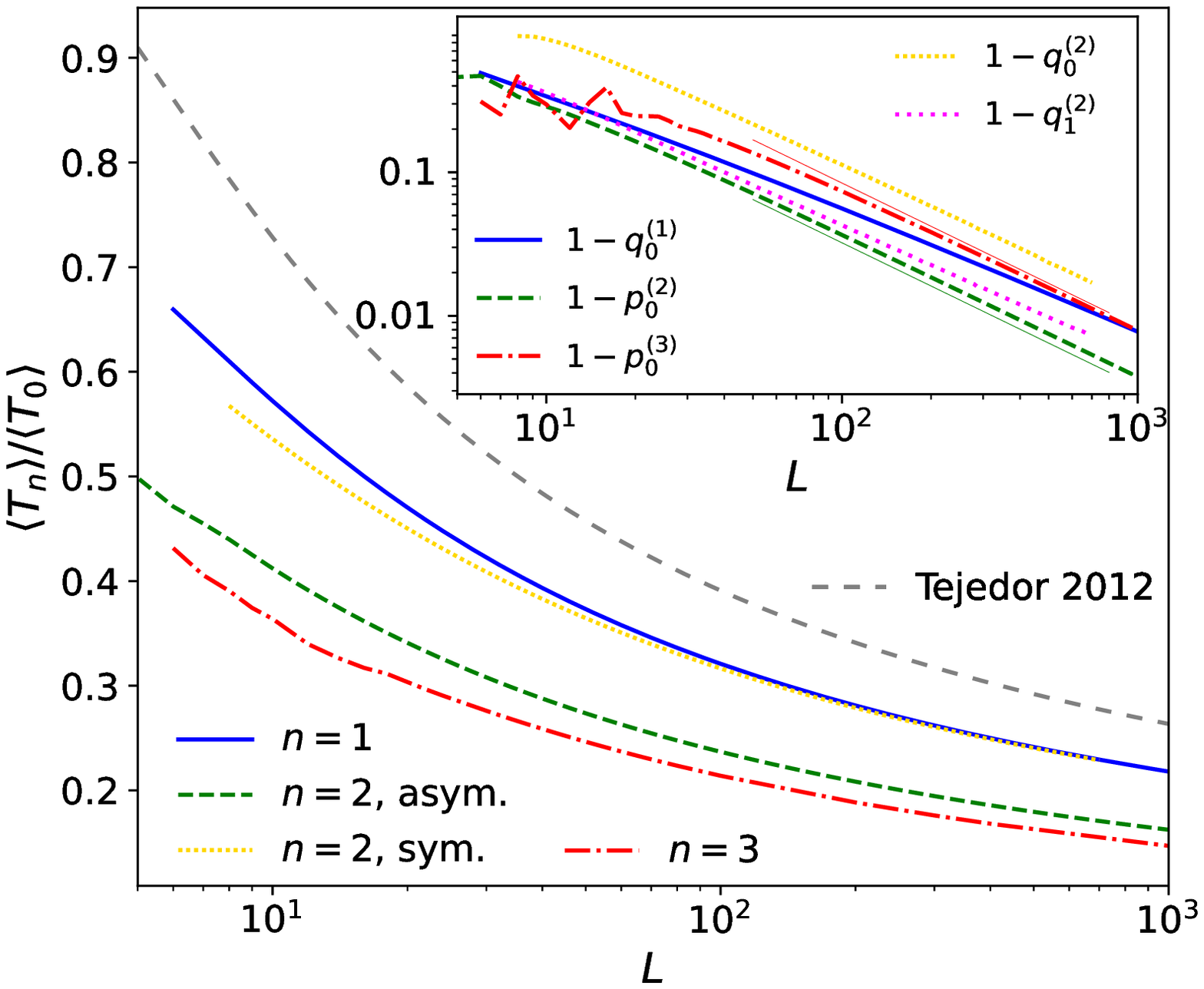}
		\end{minipage}
		\caption{Optimal search strategies on a 2-dimensional lattice for $n=1,2,3$ (left panel). In the diagrams, the sum of all arrows coming out of one box is equal to 1, therefore only the necessary coefficients are shown and all others can be deduced from the normalization constraint. For $n=2,3$, the mirror-asymmetric strategies can be decomposed into successions of two different cycles, indicated by the loops (the dashed loops correspond to the least probable cycles). The corresponding MFPT normalized by the MFPT for a blind random walk \cite{Ernst1988}, together with the optimal parameters in the inset, are shown as a function of the system size (right panel). For comparison, the optimal MFPT found by Tejedor et. al in \cite{Tejedor2012} is plotted.}
		\label{fig:optimal_walks}
	\end{center}
\end{figure*}
One intuitively expects that the number of steps $n$ kept in memory has a major impact on the search efficiency. As the case $n=0$ consists in a blind random walk, the asymptotic case $n\to\infty$ corresponds to a walk where the walker remembers all the sites it has visited and could thus elaborate a strategy to never visit twice the same site. To quantify this effect, one determines the optimal search strategy that maximizes the search efficiency for a certain value of $n$ by finding the set of conditional probabilities $p\left(\mathbf{e}_{k} | \mathbf{e}_{i_{0}}, \cdots, \mathbf{e}_{i_{n-1}} \right)$ that minimizes the MFPT. For a lattice with coordination number $z$, using the normalization constraint and assuming isotropic walks, this consists in finding the global minimum of a function of $z^{n-1}(z-1)$ variables. Using a method of coordinate descent with constraint \footnote{see appendix C of the supplemental material} for the MFPT optimization we obtain for a square lattice (z=4): 
\begin{itemize}
	\item For $n=1$, the optimal search strategy is found to be mirror-symmetric. More specifically, the probabilities $p_{l}$ and $p_r$ of turning left or right are found to be equal, while the probability of going forward is given $p_{f} = q_0^{(1)} = 1-2p_{r,l}$. The optimal 1-step memory process is therefore found to prevent from going backward. Note that $q_0^{(1)}$ depends on the system size $L$ and approaches 1 as $L\to\infty$.
	
	\item For $n=2$, the optimal strategy is mirror-asymmetric and follows the diagram shown in figure \ref{fig:optimal_walks}. Only one step in the cycle is chosen probabilistically, with probability $p_{0}^{(2)}$ that also depends on the system size. The resulting MFPT turns out to be much lower than the optimal 1-step memory process, as it is reduced by a factor $ \sim 0.75$.	
	
	If mirror-symmetry is imposed, the optimal search process is governed by two parameters, $q_0^{(2)}$ and $q_{0}^{(1)}$. However this constraint makes the MFPT almost equal to the 1-step case.
	
	\item For $n=3$, the optimal strategy is again mirror-asymmetric and governed by only one probabilistic parameter $p_{0}^{(3)}$. The corresponding diagram is shown in figure \ref{fig:optimal_walks} and the MFPT is again reduced by a factor $\sim0.86$ with respect to $n=2$. 
\end{itemize}
In all these cases, the MFPT scales proportionally to $L^2$ as $L\to\infty$, while it scales as $L^2\ln L$ for a diffusive blind random walk \cite{Condamin2007}. This explains the monotonically decreasing trend of the curves in figure \ref{fig:optimal_walks}. More precisely, it appears that $\left\langle T_1 \right\rangle \simeq L^2$, $\left\langle T_2 \right\rangle \simeq 3L^2/4$ and $\left\langle T_3 \right\rangle \simeq 2L^2/3$ as $L\to\infty$ for the optimal strategies, although the values of these prefactors are to this day not fundamentally understood. 

In addition, the dependence of the probabilities $p_0^{(n)}$ and $q_0^{(n)}$ on the system size obeys a power law of the form $p_0^{(n)} \stackrel{L\to\infty}{=} 1- aL^{-1}$ for $n>1$. The origin of this scaling and of the particular value for $a$ can be understood by decomposing the optimal search procedures for $n>1$ into 2 elementary building blocks, i.e. a preferred path $\mathcal{S}^{(n)}_{0}$ and an alternative one $\mathcal{S}^{(n)}_{+}$ (see left panel of figure \ref{fig:optimal_walks}). As the best strategy is to avoid visiting twice the same site, it is preferable to repeat the primary, longer path $\mathcal{S}^{(n)}_0$ over the entire length of the system, and then turn to the alternative one in order to avoid looping on itself. By imposing $k_c^{(n)} l_c^{(n)} = L$, where $l_c^{(n)}$ is the end-to-end distance of the path $\mathcal{S}_0^{(n)}$, and $k_c^{(n)}$ is the value of $k$ for which the probability of repeating $k$ consecutive times the path $\mathcal{S}^{(n)}_0$ is $1/2$, we obtain at first order in $1/L$ the fairly accurate estimation $p_0^{(n)} \simeq 1 - l_c^{(n)}/\ln (2) L$ (see thin full lines in the inset of figure \ref{fig:optimal_walks}, right panel).

As a comparison, the optimal MFPT found with our method is significantly  lower than the result obtained by Tejedor et al. in ref.~\cite{Tejedor2012}, where a 2D-search with a 1-step memory is considered, and the probability of going forward is favored by an amount $\epsilon$ to the 3 other directions. The gray dashed line in figure \ref{fig:optimal_walks}, also scaling as $\propto L^2$, corresponds to the value of $\epsilon$ that minimizes the MFPT in this model.

For $n>3$, the minimization procedure becomes computationally expensive, but nothing indicates that mirror-symmetric search strategies would become more favorable. In the limit of infinite memory, the optimal strategy can be simply guessed. For $n\geq L^{2}$, the walker can in fact simply scan all sites by going row by row, which is a highly mirror-asymmetric process for which the MFPT would be equal to $\left(L^2 - 1\right)/2$. Note that for $d=3$, both the MFPT for a blind walk and the optimal one with infinite memory scales proportionally to $L^3$ \cite{Condamin2007, Tejedor2012}. We therefore expect to observe the same scaling for finite values of $n$.

These results prove that memory can be useful to enhance the search efficiency of random walks. Such effects, although they might not be perfectly optimized, actually exist in some real systems, such as e.g. chemotactic walks.\newline 

\begin{figure}
	\begin{center}
		\includegraphics[width=.99\linewidth]{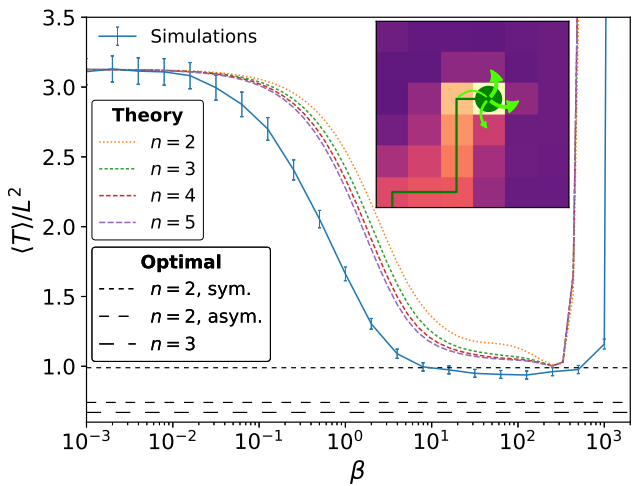}
		\caption{MFPT as a function of $\beta$ for an autochemorepulsive walk, with $D_c = 0.1$ and $L=100$, from simulation (full line with error bars were computed via Jackknife re-sampling \cite{Efron1981}) and theory for various values of $n$ (dotted lines). Inset: sketch of the auto-chemorepulsive searcher. Size of the arrows correspond to $p_{i\rightarrow j}$, color code to concentration values of $c_j$.}
		\label{fig:mfpt_tsaw}
	\end{center}
\end{figure}

Next we consider the auto-chemotactic searcher, and focus on chemo-repulsive searcher-cue interactions since they avoid repeated visits of already scanned areas and thereby increasing the search efficiency. A simple lattice model for an autochemorepulsive walk can be constructed as an adaptation of the {\it true self-avoiding walk} with chemical diffusion \cite{Amit1983, Grassberger2017}. A concentration field $c$ is defined on the lattice, and diffuses at each time step according to a discretized diffusion equation, with diffusion constant $D_c$. After the diffusive step, the searcher moves from site $i$ to site $j$ with probability  
\begin{equation}
	\label{eqn:proba_TSAW}
	p_{i\to j} = \left[ 1 + \sum_{k\neq j} \exp \left(-\beta \left( c_k - c_j \right) \right) \right]^{-1}
\end{equation}
where the sum runs over all neighbors of $i$, except $j$. Here, $\beta$ quantifies the coupling between the walker and the concentration field : for $\beta\to 0$, the process reduces to a unbiased blind random walk, while the limit $\beta\to \infty$ corresponds to the case where the walker always jumps to the neighboring site with the lowest concentration. Finally, once the walker has jumped to the site $j$, it adds an amount $\delta c$ to the concentration field at this site. 

Because the profile of the concentration field at a certain time depends on the entire path of the walker, this model is obviously a non-Markovian process. We determine the conditional probabilities at time after n steps, $p\left(\mathbf{e}_{k} | \mathbf{e}_{i_{0}}, \cdots, \mathbf{e}_{i_{n-1}}; t=n \right)$, starting from a zero concentration field, $c_i=0$ on all sites $i$. These probabilities for all possible $n$-step paths are then used as inputs for the formalism introduced in the previous paragraph and the MFPT can therefore be estimated. This approach is obviously more accurate for larger values of $n$ but the exponential computational cost forbids to implement it for very large values. Still, relatively low values of $n$ can predict the qualitative behavior of the MFPT.

Figure \ref{fig:mfpt_tsaw} shows the mean first-passage time of the autochemorepulsive walk for a 2-dimensional lattice of size $L=100$, using the formalism presented in this paper, together with simulation results (each point accounts for $10^4$ trajectories)\footnote{An animation of a simulated trajectory is available as a supplemental material}. From both theory and simulations, it appears clearly that for a certain value of $D_c$, there exists an optimal value for $\beta$ that minimizes the search time. At low values of $\beta$ the MFPT slowly decreases as the process goes from a blind random walk to a smarter walk in which the chemical information from the environment is used. However, as $\beta$ gets larger, the MFPT abruptly increases. This effect can be understood as follows : after the walker has jumped to a certain site, and because it has released some cue behind it, the chemical concentration is expected to be lower on the forward site than on the left and right sites, and even more than on the backward site. For large values of $\beta$, as the walker jumps on the neighboring site with the lowest concentration with probability $p\sim1$, it will thus tend to go forward, and so over very long distances, turning its motion into an almost ballistic behavior. Fully ballistic trajectories are obviously not efficient for a search process, which we observe  here with the very large values for the MFPT for $\beta\to\infty$. This transition from diffusive to ballistic behavior can be quantified by the persistence length $l_p$, defined as the mean number of consecutive steps in the same direction, and which is shown in figure \ref{fig:lp_tsaw} (numerical vs. analytical estimates \footnote{see Appendix D of the supplemental material}.) It confirms that the walker's persistence length strongly increases for large values $\beta$. To optimize its search, an autochemotactic particle must find the right balance between a blind search that makes use of no chemical information and a strong coupling with the cue that makes it go in a straight line. 
\begin{figure}
	\begin{center}
		\includegraphics[width=.99\linewidth]{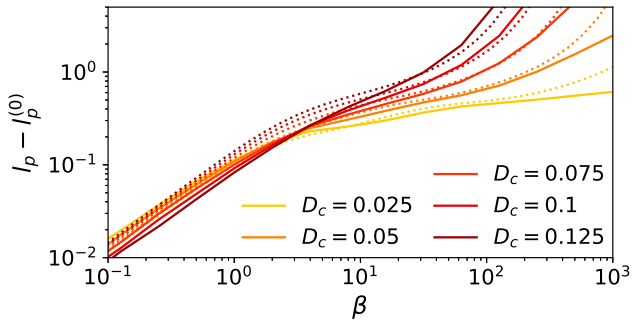}
		\caption{Persistence length for the autochemotactic walk as a function of $\beta$. The limit value $l_{p}^{(0)} = 4/3$ is subtracted. Discrepancies between theory (dotted lines) and simulations (full lines) as $\beta\to\infty$ are mostly finite-size effects.}
		\label{fig:lp_tsaw}
	\end{center}
\end{figure}

In order to compare the optimal strategy of the autochemotactic search with the optimal	n-step search strategies we discussed in the first part, we compute the conditional probabilities  $p\left(\mathbf{e}_k|\mathbf{e}_{i_0}, \cdots, \mathbf{e}_{i_{n-1}}\right)$ of the optimal autochemotactic search in the stationary regime as follows: the number of steps in a particular direction $\mathbf{e}_k$ following a particular $n$-step path $\left\{\mathbf{e}_{i_0}, \cdots, \mathbf{e}_{i_{n-1}}\right\}$ is sampled, and its average over the 4 possible values of $\mathbf{e}_k$ yields the corresponding transitional probability. Note that these stationary probabilities will be different from the probabilities computed for our theoretical estimation, where the $n$-step path was initialized with a zero concentration field. Also note that the auto-chemotactic  walk is not fully described by the $n$-step conditional probabilities $p\left(\mathbf{e}_k|\mathbf{e}_{i_0}, \cdots, \mathbf{e}_{i_{n-1}}\right)$, but by an infinite hierarchy of conditional probabilities, which we truncate after n steps. 
	
For the optimal point $\left\{D_c=0.1, L=100,  \beta=0.01\right\}$, this analysis performed with $n=2,3$ shows that the autochemotactic search for the optimal value of $\beta$ does not mimic the optimal search strategies of figure \ref{fig:optimal_walks} \footnote{see Appendix E of the Supplemental Material}. A few major differences can be noticed. First, while the optimal strategies allow only very few transitional probabilities to be different from 0 and 1, the best autochemotactic search is intrinsically more random as more of those quantities have intermediate values. Second, the autochemotactic walk is {\it by definition} mirror-symmetric. Finally, the optimal strategy for $n=3$ allows the walker to turn in the same direction twice in a row, but this move is never observed in the best autochemotactic search.  As a comparison, we show in figure \ref{fig:mfpt_tsaw} the value of the MFPT obtained using the optimal $n$-step strategies presented in the previous paragraphs, for $n=2,3$. For $D_c=0.1$, the optimal value for $\beta$ in the autochemotactic does not beat the optimal asymmetric strategies, but it does result in a slightly lower MFPT than the optimal symmetric strategy with a 2-step memory. The autochemotactic walk can therefore still be considered as an efficient search strategy.
\newline

\vspace*{2ex}
The results presented in this letter clearly indicate that non-Markovian features of search processes can be tuned in order to maximize search efficiency. Optimal search strategies are found to be mirror-asymmetric and more efficient with longer memory. However, search processes in nature are not necessarily as optimal, but the physical parameters that govern them can still be adjusted to improve efficiency. The formal and systematic tool introduced in this paper should be useful for other biologically relevant applications, many of which present non-negligible non-Markovian effects.

\subsection*{Acknowledgements}
We thank Raphaël Bulle and Dr. Adam Wisocki for stimulating discussions and acknowledge financial support by the DFG via the Collaborative Research Center SFB 1027.

\newpage
\clearpage

\onecolumngrid

\subsection*{Appendix A - Matrix structures}

We show the structure of the matrix $\mathbf{P}_n$, that is at the core of our formalism. As explained in the main text, we first need to cast all possible paths $\left\{\mathbf{e}_{i_{0}}, \cdots, \mathbf{e}_{i_{n-1}} \right\}$ into a vector $\mathbf{s}$ of size $z^n$. The most intuitive way to perform this operation is to vary hierarchically the directions in the path as one go from one entry of the vector to the next. Noting the unit vectors $\mathbf{e}_{0}, \cdots, \mathbf{e}_{z-1}$, a natural sorting would be as follows
\begin{align*}
s_{0} =& \left\{\mathbf{e}_{0}, \mathbf{e}_{0}, \cdots, \mathbf{e}_{0}, \mathbf{e}_{0} \right\} \\
s_{1} =& \left\{\mathbf{e}_{0}, \mathbf{e}_{0}, \cdots, \mathbf{e}_{0}, \mathbf{e}_{1} \right\} \\
\cdots&  \\
s_{z-1} =& \left\{\mathbf{e}_{0}, \mathbf{e}_{0}, \cdots, \mathbf{e}_{0}, \mathbf{e}_{z-1} \right\} \\
s_{z} =& \left\{\mathbf{e}_{0}, \mathbf{e}_{0}, \cdots, \mathbf{e}_{1}, \mathbf{e}_{0} \right\} \\
s_{z+1} =& \left\{\mathbf{e}_{0}, \mathbf{e}_{0}, \cdots, \mathbf{e}_{1}, \mathbf{e}_{1} \right\} \\
\cdots&  \\
s_{z^n-2} =& \left\{\mathbf{e}_{z-1}, \mathbf{e}_{z-1}, \cdots, \mathbf{e}_{z-1}, \mathbf{e}_{z-2} \right\} \\
s_{z^n-1} =& \left\{\mathbf{e}_{z-1}, \mathbf{e}_{z-1}, \cdots, \mathbf{e}_{z-1}, \mathbf{e}_{z-1} \right\} 
\end{align*} 
Using this ordering, the matrix $\mathbf{P}_n$ has the structure
\[ 
\mathbf{P}_n = \left(
\begin{tabular}{c}
$\mathbf{P}_{n}^{(0)}$ \\
$\vdots$ \\
$\mathbf{P}_{n}^{(z^{n-1}-1)}$
\end{tabular}
\right) 
\] 
where each submatrix $\mathbf{P}_n^{(i)}$ is of size $z\times z^{n}$ and is of the form
\[ 
\mathbf{P}_n^{(i)} = 
\left(
\begin{tabular}{cccc}
$\mathbf{P}_n^{(i,0)}$ & 0                      & $\cdots$ & 0 \\ 
0                      & $\mathbf{P}_n^{(i,1)}$ & $\ddots$ & $\vdots$ \\
$\vdots$               & $\ddots$               & $\ddots$ & 0 \\ 
0                      & $\cdots$               & 0        & $\mathbf{P}_n^{(i,z-1)}$
\end{tabular}
\right) 
\] 
Here, each matrix $\mathbf{P}_n^{(i,j)}$ is a row matrix with $z$ columns, namely 
\[
\mathbf{P}_n^{(i,j)} = \left(p(\mathbf{e}_0|s_{\alpha(i,j)}) \ \ \cdots \ \ p(\mathbf{e}_{z-1}|s_{\alpha(i,j)})   \right)
\]
where $\alpha(i,j) = iz+j$. The non-zero elements of these matrices can be gathered into matrices $\mathbf{Q}_n^{(i)}$ of the form
\[ 
\mathbf{Q}_n^{(i)} = \left(
\begin{tabular}{c}
$\mathbf{P}_{n}^{(i,0)}$ \\
$\vdots$ \\
$\mathbf{P}_{n}^{(i,z-1)}$
\end{tabular}
\right) 
\] 
For isotropic walks, all matrix $\mathbf{Q}_n^{(i)}$ contain exactly the same elements in different orders such that the matrix $\mathbf{Q}_n^{(0)}$ is the building block of the formalism. More precisely, the matrix $\mathbf{Q}_n^{(i)}$ contains all probabilities $p(\mathbf{e}_{k}|\mathbf{e}_{i_0},\cdots,\mathbf{e}_{i_{n-1}})$ with $i_0=i$. To draw a connection between different values of $i$, consider $\mathbf{R}$ a transformation that maps the lattice onto itself. It holds
\begin{equation}
p(\mathbf{e}_{k}|\mathbf{e}_{i},\cdots,\mathbf{e}_{i_{n-1}}) =	p(\mathbf{R}\mathbf{e}_{k}|\mathbf{R}\mathbf{e}_{i_0},\cdots,\mathbf{R}\mathbf{e}_{i_{n-1}})
\end{equation}
Now, let $\mathbf{R}_{j}$ be a transformation such that $\mathbf{e}_{j} = \mathbf{R}_{j}\mathbf{e}_0$ : 
\begin{align}
p(\mathbf{e}_{k}|\mathbf{e}_{i}, &\mathbf{e}_{i_1}, \cdots,\mathbf{e}_{i_{n-1}}) \nonumber \\
&=	p(\mathbf{R}_{-i}\mathbf{e}_{k}|\mathbf{e}_{0}, \mathbf{R}_{-i} \mathbf{e}_{i_1} ,\cdots,\mathbf{R}_{-i_0}\mathbf{e}_{i_{n-1}})
\end{align}
This identity allows to relate the elements of $\mathbf{Q}_{n}^{(i)}$ to those of $\mathbf{Q}_{n}^{(0)}$.

As an example, consider a 2-dimensional square lattice. There are $z=4$ unit vectors $\mathbf{e}_k$ (with the convention $\mathbf{e}_{k+z} = \mathbf{e}_{k} $), and the transformations $\mathbf{R}_{j}$ can be defined as $\mathbf{R}_{j} = \mathbf{R}_0^{j}$ where $\mathbf{R}_0$ is a 90°-rotation. Moreover, we have $\mathbf{R}_{j}\mathbf{e}_{k}=\mathbf{e}_{k+j}$. This yields 
\begin{equation}
p(\mathbf{e}_{k}|\mathbf{e}_{i}, \mathbf{e}_{i_1}, \cdots,\mathbf{e}_{i_{n-1}}) =	p(\mathbf{e}_{k-i}|\mathbf{e}_{0}, \mathbf{e}_{i_1-i} ,\cdots,\mathbf{e}_{i_{n-1}-i})
\end{equation}
Defining $\sigma'_{i} = \sum_{j=1}^{n-1} \left[(i_{j}-i)\mod z\right]z^{n-j-2}$, this identity can be transformed to relate matrix elements of $\mathbf{Q}_{n}^{(i)}$ to those of $\mathbf{Q}_{n}^{(0)}$, namely 
\[
{Q_{n}^{(i)}}_{\sigma'_{0},k} = {Q_{n}^{(0)}}_{\sigma'_{i},(k-i)\!\!\!\!\!\mod z}
\]
Finally, note that because the matrix $\mathbf{E}(\mathbf{q})$ is a diagonal matrix, the structure of the matrix $\mathbf{P}_0\mathbf{E}(\mathbf{q})$ will be similar as the one of $\mathbf{P}_0$.

\subsection*{Appendix B - Computational cost}
In the formalism proposed in this paper, the MFPT is found by a matrix inversion. In particular, one inversion must be performed per each possible value of $\mathbf{q}$. Given that the computational cost of the inversion of a matrix of size $n\times n$ is proportional to $n^{\alpha}$, the scaling of our computational method should as $t_{CPU}\sim L^d z^{\alpha n}$. For large values of $n$, this might become particularly expensive. This can be dealt with by rewriting the problem in a different way and using an approximate method for the matrix inversion. 

As explained in the main text, the MFPT can be written as a dot product $\mathbf{p}_{n}\cdot\left\langle \mathbf{t}_{n} \right\rangle$, with $\mathbf{p}_n = \mathbf{M}_n^{-1}\mathbf{v}_n$ and $\left\langle \mathbf{t}_n \right\rangle = \sum_{\mathbf{q}\neq 0} \left(\mathbb{I} - \mathbf{P}_n\mathbf{E}_n(\mathbf{q})\right)^{-1}\mathbf{u}_n$. Let us now write this as 
\begin{align*}
\left\langle T_n \right\rangle =& \sum_{i} {p_n}_{i} \left\langle {t_n}_i \right\rangle \\
=& \sum_{\mathbf{q}\neq 0} \sum_{i} \sum_{j,k}  \left[\left(\mathbb{I} - \mathbf{P}_n\mathbf{E}_n(\mathbf{q})\right)^{-1}\right]_{ji} \left(\mathbf{M}_n^{-1}\right)_{ki} {u_n}_{j} {v_n}_{k} \\
=& \sum_{\mathbf{q}\neq 0}  \sum_{j,k}  {u_n}_{j} {v_n}_{k} \sum_{i} \left[\left(\mathbb{I} - \mathbf{P}_n\mathbf{E}_n(\mathbf{q})\right)^{-1}\right]_{ji} \left(\left({\mathbf{M}^{T}_n}\right)^{-1}\right)_{ik}  
\end{align*}
By noting $\mathbf{H}_n(\mathbf{q}) = \mathbf{M}_n^{T} \left(\mathbb{I} - \mathbf{P}_n\mathbf{E}_n(\mathbf{q})\right)$ we obtain
\begin{align*}
\left\langle T_n \right\rangle =& \sum_{\mathbf{q}\neq 0}  \sum_{j,k} {u_n}_{j} {v_n}_{k} \left(\mathbf{H}_n(\mathbf{q})^{-1}\right)_{jk}  
\end{align*}
Now, we use ${u_n}_{j} = 1$ for any $j$, and ${v_n}_{k} = \delta_{k,z^{n}-1}$ and find
\begin{align*}
\left\langle T_n \right\rangle =& \sum_{\mathbf{q}\neq 0}  \sum_{j} \left(\mathbf{H}_n(\mathbf{q})^{-1}\right)_{j,z^{n}-1}  
\end{align*}
The MFPT is therefore equal to the sum of the elements of the last column of the inverse of $\mathbf{H}_n$. If the matrix size $z^n$ is not too large, the matrix can be entirely inverted without much effort. However, this procedure becomes very expensive for large matrices. As one only needs one column of the matrix $\mathbf{H}_n^{-1}$, one could use approximate methods to estimate specific matrix elements. This is what Bai et al. proposed in ref. \cite{Bai1996}, in which the authors provide an algorithm to come up with upper an lower bounds for specific elements of matrix inverses without having to invert the matrix entirely.

\subsection*{Appendix C - Minimization algorithm}

In order to find the optimal algorithm for a $n$-step search process, we have a applied a coordinate-descent minimization procedure with constraint. In our formalism, the MFPT $T$ is a function of the conditional probabilities $p\left(\mathbf{e}_{k} | \mathbf{e}_{i_{0}}, \cdots, \mathbf{e}_{i_{n-1}} \right)$. Because of rotational symmetry, there are $N=z^{n}$ different such coefficients. In addition, it holds $\sum_{k=0}^{z-1} p\left(\mathbf{e}_{k} | \mathbf{e}_{i_{0}}, \cdots, \mathbf{e}_{i_{n-1}} \right) = 1$, which consists in a set of constraints. Let us now order all these coefficients in a vector $\mathbf{x}=\left\{x_0, \cdots, x_{N-1}\right\}$, and minimize the function $T(\mathbf{x})$ following algorithm \ref{algo_min}. This procedure is repeated many times, always starting from a different initial guess in order to distinguish local minima from the global minimum.
\begin{figure}[h]
	\label{algo_min}
	\fontsize{10}{15}\selectfont	
	\begin{algorithmic}
		\STATE{Initialize $\mathbf{x}$ randomly, ensuring $\forall j\in[\![0, N-1 ]\!], \ \sum_{i=jz}^{(j+1)z-1}x_{i} = 1$;}
		\WHILE{$\Delta T>\epsilon$}
		\STATE{$\chi \leftarrow $ random permutation of $[\![0, N-1 ]\!]$;}
		\FOR{$i \in [\![0, N-1 ]\!]$}
		\STATE{$j\leftarrow \chi_{i}$;}
		\STATE{$k_0 \leftarrow z\left\lfloor \frac{j}{z} \right\rfloor$;}
		\STATE{choose randomly $k \in \left[\!\!\left[ k_0; k_0 + z \right]\!\!\right] \setminus \{j\}$;}
		\STATE{$s\leftarrow \sum_{k'=k_0}^{k_0 + z}x_{k'} - x_j - x_k$;}
		\STATE{$x_{j} \leftarrow \underset{y}{\argmin}\ T(x_0,\cdots,x_{j-1}, y, x_{j+1}, \cdots , x_{k-1}, 1-y-s, x_{k+1}, \cdots, x_{N-1})$;}
		\STATE{$\Delta T = |T_0 - T(\mathbf{x})|$;}
		\STATE{$T_{0}\leftarrow T(\mathbf{x})$;}
		\ENDFOR	
		\ENDWHILE		
	\end{algorithmic}
	\captionsetup{labelformat=empty}
	\caption{\underline{Algorithm 1} : Random-order coordinate-descent with constraint}
\end{figure}

\subsection*{Appendix D - Evaluation of the persistence length in the autochemotactic model}

In the autochemotactic model, the persistence length $l_p$ is defined as the mean number of consecutive steps in the same direction. By noting $p_n$ the probability of continuing forward after $n$ steps in the same direction, it holds
\begin{equation}
l_{p} =  \sum_{n=1}^{\infty} n (1-p_{n}) \prod_{k=1}^{n-1} p_{k}
\end{equation}
Now, we need to quantify $p_n$ in terms of the parameters of the model. Consider a walker that has changed its direction at time $t_0$. In principle, $p_n$ accounts for the concentration field $c^{(0)}(n)$ resulting from the diffusion of the concentration profile at time $t_0$ and for the newly deposited chemical along the $n$ steps forward up to time $t_0+n$, resulting into a chemical field $c^{(+)}(n)$ :
\begin{equation*}
p_{n} =  \exp\left(-\beta (c_i^{(0)}(n)+c_i^{(+)}(n))\right) / \sum_{j}\exp\left(-\beta (c_j^{(0)}(n)+c_j^{(+)}(n))\right) 
\end{equation*}
where the site $i$ is the forward site w.r.t the site $k$ reached after $n$ steps in the same direction, and the sum runs over all neighboring sites of $k$. 

Here, we decide to neglect the contribution of $c^{(0)}(n)$ and focus only on the contribution of $c^{(+)}(n)$. This corresponds to considering that the chemical field resets to zero before each turn the walker takes. This is of course a rough approximation but it is qualitatively good enough to quantify the long-persistence regime. The concentration field $c^{(+)}(n)$ can be computed exactly and is of the form
\begin{equation*}
c^{(+)}_i(n) = \sum_{l=0}^{n+1} \alpha_{i,l} D_c^{l}
\end{equation*}
In the low-diffusion limit $D_c\ll 1$, we truncate this sum at order $2$ and obtain 
\begin{equation*}
p_n^{-1} = \left\{
\begin{tabular}{l l}
$1 + e^{-\beta}  + e^{-3\beta D_{c}}  + e^{-\beta D_{c}^{2}} $ & ifr $n=1$ \\
$1 + e^{-\beta}  + e^{-7\beta D_{c}^{2}}  + e^{-\beta D_{c}^{2}}$ & if $n=2$ \\
$1 + e^{-\beta}  + 2e^{-\beta D_{c}^{2}}$ & for $n>2$
\end{tabular}
\right.
\end{equation*}
The constant value of $p_n$ for $n>2$ finally allows to write the persistence length as 
\begin{align}
l_{p} =& (1 - p_{1}) + 2p_{1}(1 - p_{2}) + p_{1}p_{2} \sum_{n=3}^{\infty} np_{3}^{n-3}(1-p_{3}) \nonumber \\
=& 1 + p_{1} - 2p_{1}p_{2} + p_{1}p_{2}\frac{1-p_3}{p_3} \sum_{n=3}^{\infty} np_{3}^{n-2}\nonumber \\
=& 1 + p_{1} - 2p_{1}p_{2} + p_{1}p_{2}\frac{1-p_3}{p_3} \left[(1-p_3)^{-2} + (1-p_3)^{-1} - 2\right]\nonumber \\
l_p=& 1 + p_{1} \left(1 + \frac{p_{2}}{1-p_3}  \right)
\end{align}

\subsection*{Appendix E - Effective transitional probabilities of the autochemotactic walk}

As mentioned in the main text, the transitional probabilities $p\left(\mathbf{e}_k|\mathbf{e}_{i_0}, \cdots, \mathbf{e}_{i_{n-1}}\right)$ can be effectively computed from the simulation data of the autochemotactic walk. We present here the matrix $\mathbf{Q}_n^{(0)}$ introduced in Appendix A, for $n=2,3$, for the autochemotactic walk, and the optimal ones obtained by minimizing the MFPT. 

\begin{equation*}
\text{\underline{Optimal strategy}:} \ \ \mathbf{Q}_2^{(0)} = \left(
\begin{tabular}{cccc}
0 & 1 & 0 & 0\\
1 & 0 & 0 & 0\\
0 & 0 & 1 & 0\\
$1-p_0^{(2)}$ & 0 & 0 & $p_0^{(2)}$
\end{tabular}
\right) , \hspace{2.5ex}
\let\scriptstyle\textstyle \substack{ \text{\underline{Autochemotactic walk}:} \\ D_c=0.1, \ L=100 \\ \beta=100 }\ \ \mathbf{Q}_2^{(0)} = \left(
\begin{tabular}{cccc}
0.68 & 0.16 & 0 & 0.16 \\
0.71 & 0 & 0 & 0.29 \\
$\bullet$ & $\bullet$ & $\bullet$ & $\bullet$ \\
0.71 & 0.29 & 0 & 0 \\
\end{tabular}
\right) 
\end{equation*}

\begin{equation*}
\text{\underline{Optimal strategy}:} \ \ \mathbf{Q}_3^{(0)} = \left(
\begin{tabular}{cccc}
0 & 0 & 0 & 1\\
0 & $p_0^{(3)}$ & $1-p_0^{(3)}$ & 1\\
0 & 1 & 0 & 0\\
1 & 0 & 0 & 0\\
0 & 0 & 0 & 1\\
0 & 1 & 0 & 0\\
0 & 1 & 0 & 0\\
0 & 0 & 1 & 0\\
0 & 1 & 0 & 0\\
1 & 0 & 0 & 0\\
0 & 0 & 1 & 0\\
0 & 0 & 1 & 0\\
0 & 1 & 0 & 0\\
1 & 0 & 0 & 0\\
0 & 0 & 1 & 0\\
1 & 0 & 0 & 0
\end{tabular}
\right) , \hspace{2.5ex}
\let\scriptstyle\textstyle \substack{ \text{\underline{Autochemotactic walk}:} \\ D_c=0.1, \ L=100 \\ \beta=100 }\ \ \mathbf{Q}_3^{(0)} = \left(
\begin{tabular}{cccc}
0.64 & 0.18 & 0 & 0.18 \\ 
0.70 & 0 & 0 & 0.30 \\ 
$\bullet$ & $\bullet$ & $\bullet$ & $\bullet$ \\
0.70 & 0.30 & 0 & 0 \\ 
0.74 & 0.02 & 0 & 0.24 \\ 
$\bullet$ & $\bullet$ & $\bullet$ & $\bullet$ \\
$\bullet$ & $\bullet$ & $\bullet$ & $\bullet$ \\
0.73 & 0.27 & 0 & 0 \\ 
$\bullet$ & $\bullet$ & $\bullet$ & $\bullet$ \\
$\bullet$ & $\bullet$ & $\bullet$ & $\bullet$ \\
$\bullet$ & $\bullet$ & $\bullet$ & $\bullet$ \\
$\bullet$ & $\bullet$ & $\bullet$ & $\bullet$ \\ 
0.74 & 0.24 & 0 & 0.02 \\ 
0.73 & 0 & 0 & 0.27 \\ 
$\bullet$ & $\bullet$ & $\bullet$ & $\bullet$ \\
$\bullet$ & $\bullet$ & $\bullet$ & $\bullet$ \\
\end{tabular}
\right) 
\end{equation*}
In the case of the autochemotactic walks, the bullets indicate that none of the $n$-step paths of an entire line were observed in the simulations, such that no normalization was possible. Still, we clearly see that the search strategy of the autochemotactic walker is very different from the optimal ones.

\end{document}